\documentstyle[12pt]{article}
\begin{document}
\baselineskip = 20pt
\title{\bf Notes On The Born-Oppenheimer Approach In
A Closed Dynamical System }

\author{{\bf Dhurjati Prasad Datta
}\\
 Department of Mathematics\\
North Eastern Regional Institute of Science and Technology\\
Itanagar-791109, Arunachal Pradesh, India\thanks {Permanent address}\\
{\normalsize and}\\
International Centre for Theoretical Physics\\
Post Bag 586, 34100 Trieste, Italy\\ \\
email:dpd@nerist.ernet.in}

\date{}
\maketitle

\begin{abstract}

The  various  recent  studies  on  the  application  of   the
Born-Oppenheimer approach in a closed  gravity  matter  system  is
examined. It is pointed out that the Born-Oppenheimer approach  in
the absence of an a priori time is likely to yield potentially new
results.
\end{abstract}
\newpage
\par
Recently, there   has   been    a    renewed    interest    in
the Born-Oppenheimer  (BO)  approach  in  analyzing   the   quantum
evolution  of  composite  systems  involving  two  separate   mass
(energy) scales [1-8]. It is now well known that  the  conventional
BO   approximation   method   in   atomic/molecular    physics, for
instance, not only offers a  natural  framework  in  realizing  the
Berry adiabatic phase in the quantum state of the lighter system,
but the motion of the heavy system also  gets  influenced  by  the
induced geometric gauge potentials, besides the usual BO potential
determined  by  the  energy  expectation  value  of  the   lighter
state[1]. The effects  of  the  geometric  electric  and  magnetic
fields on the heavy system have been shown to yield high degree of
agreements with the exact results[2]. Moreover,the  predictions  of
the BO  analysis  in  localized  quantum  mechanical  systems  are
unambiguous and  well  tested  in  laboratory  (atomic/  molecular
physics) experiments[1].
\par
Applications of the BO  analysis  in  closed  gravity-matter
systems in quantum cosmology, on the other hand, seem still to  be
rather inconclusive. Although the plausibility of an application of
a BO type analysis in a quantized gravity matter  system  is  well
accepted ( because of the large value of the Placnk mass  compared
to the mass scales of ordinary  matter)[3], there  seems  to  be  a
considerable amount of disagreement in  interpreting  the  results
obtained in this method[3-8]. The main reason of these disparities
of course relates to the issue  of  time  in  quantum  gravity.  A
gravity matter system, for instance,  in  a  quantum  cosmological
model,is a closed dynamical system in the sense that there  is  no
interaction external to those induced by  the  mutual  (and  self)
interactions between the given gravitational and matter degrees of
freedom. The  corresponding   quantum   dynamical   equation-   the
so-called Wheeler -Dewitt (WD)  equation-  which  is  obtained  by
applying the Dirac quantization rule to the classical  Hamiltonian
constraint in a generally covariant theory  of  gravity,thus  does
not involve any  external  concept  of  time. This  leads  to  deep
conceptual as  well  as  interpretational  problems  in  canonical
quantum gravity.A basic motivation in the  semiclassical  approach
[3] in quantum gravity is to understand  how  a  concept  of  time
could emerge intrinsically from  an  apparently  timeless  quantum
'evolution' equation in a quasi-classical regime of the  (massive)
gravitational degrees of freedom. Although a semiclassical approach
does not aim at solving the issue of time in quantum gravity,  the
study is likely to offer new insights into the nature of time  and
related issues in canonical quantum gravity.
\par
Recent studies in the BO type analysis in  quantum  cosmology
seem to appreciate the importance of incorporating  the  geometric
phases and the associated gauge potentials in the  discussions  of
the effective dynamics of the  gravitational  degrees  of  freedom
[3-8]. However, as  remarked   already, the   descriptions   of   the
effective dynamics seem to vary because  of  an  arbitrariness  in
obtaining the back reaction of matter on gravity, which in fact is
related to the  the  definition  of  the  intrinsic  time. In  this
paper, we analyze the origin of  this  ambiguity  in  realizing  an
intrinsic time in the BO framework  and  point  out  that  the BO
analysis of a composite system in the absence of an external  time
would  lead  to  nontrivial  predictions[8] in  comparison  to   the
conventional treatments.
\par
Let us  consider  an  interacting  system  described  by  the
Hamiltonian
\par
$$
H(q,\phi ) = {1\over 2M} G^{ij}P_{i}P_{j} + MV(q) + H_{m}(\phi ,q)
\eqno(1)$$
\noindent Here,$G^{ij}$  denote  the  metric  tensor  in  the  space   
of   heavy
configuration variables $q_{i}(i=1,2,3,..N), P_{i}$being the corresponding
conjugate momentum. $H_{m}$ is the Hamiltonian  of  the  lighter  system
$\phi $, which depends parametrically on the heavy variables q. The  mass
$m$ of the lighter system is  much  less  than  that  of  the  heavy
system, $m<<$M. The composite Hamiltonian (1) can be considered as  a
model of the minisuperspace  quantum  cosmological  gravity-matter
system provided $G^{ij}$ is  interpreted  as  the  superspace  (Dewitt)
metric  of  compact  three  geometries $q$  and $\phi $  is  the  matter
fields. The mass $M$ then stands for the Planck  mass  squared. For  a
molecular system, on the  other  hand, $G^{ij}=\delta ^{ij}$  and $M$  denotes  the
molecular mass. The signature of the metric, however, is  not  very
crucial for our present formal discussion.
\par
To identify the actual source of ambiguity in the BO  method
in quantum cosmology and for the sake of generality, we  begin  our
analysis in the framework of the ordinary quantum  mechanics  with
an {\it a priori} time. The quantum evolution of the composite system (1)
is then governed by the Schr\"odinger equation
\par
$$
H(q,\phi )\Psi (q,\phi )= E\Psi (q,\phi ) =
i{\partial \over \partial t} \Psi (q,\phi )
\eqno(2)$$
\noindent $\Psi $ is thus an energy eigenstate of the total Hamiltonian:
$\Psi =e^{-\hbox{iEt}}\Psi _{0}$.
Assuming  that $H(q,\phi )$  is  not   explicitly $t$   dependent   and
constraining $\Psi $ to the zero energy state (which in fact amounts  to
a redefinition of the potential $V \rightarrow V-E/M$), the eigenstate  
equation (2) mimics the WD equation
\par
$$
H\Psi _{0}=0
\eqno(3)$$
\noindent One can thus proceed to study the effective dynamics of the  heavy
$q$-system under the influence of the lighter system $\phi $, both with or
without an  external  time, in  the  present  model.  Although  the
mathematical framework is more or less similar,the two  situations
are clearly distinct physically; thus necessitating an extra  care
in interpreting and comparing the  relevant  results. To  keep  our
discussions sufficiently general we thus choose to work  with  the
more general equation,viz. eq.(2)
\par
$$
({1\over 2M}G^{ij}P_{i}P_{j} +MV(q) +H_{m})\Psi _{0} =E\Psi _{0}
\eqno(4)$$
\noindent where $P_{i}=-i\hbar \partial /\partial q^{i}$, in the following. 
The total energy $E$ here is  a
constant(which may even be zero), reflecting the constraint nature
of the operator eq.(4).
\par
We make a BO decomposition of $\Psi $ as
\par
$$
\Psi _{0}(q,\phi )=\psi (q)\chi (q,\phi )
\eqno(5)$$
\noindent where the quantum state of the lighter system $\chi (q,\phi )$ is not
further decomposable. Projecting the  total  equation  (4)  on  the
normalized lighter state $\chi  (<\chi \mid \chi >=1)$ one gets,

\par
$$
\left[-{\hbar ^{2}\over {2M}}G^{ij}D_{i}D_{j} +MV+ 
<H_{m}>+{1 \over{2M}} \rho \right]\psi (q)=0\eqno(6)
$$
\noindent where $<H_{m}>=<\chi \mid H_{m}\mid \chi >$ and the 
covariant derivatives $D_{i}=\partial /\partial q^{i}+iA_{i}$ and
$\bar{D}_{i}=\partial /\partial q^{i}-iA_{i}$ are introduced because  
of  the  induced  (adiabatic) $U(1)$ magnetic connection
\par
$$
A_{i}=-i<\chi \mid \partial /\partial q^{i}\mid \chi >
\eqno(7)$$
\noindent and $\rho$ is the electric potential
\par
$$                \rho = - \hbar^{2}<\chi\mid G^{ij}\bar D_{i}\bar D_{j} \mid \chi>
=-\hbar^{2}<\bar D^{2}>
\eqno(8)$$
\par
Further,on multiplying eq.(6) by $\chi $ and  subtracting  it  from
eq.(4) one obtains following ref.[4-6]
\par
$$
\left(H_{m}-<H_{m}>\right)\chi -{\hbar ^{2}\over M}\psi ^{-1}
G^{ij}(D_{i}\psi )\bar{D}_{j}\chi ={\hbar^{2} \over {2M}}
\left(\bar D^{-2}- <\bar{D}^{2}>\right)\chi \eqno(9)
$$
\noindent Note that eq.(6) and (9) constitute an exact  set of coupled  (non
linear ) equations and should be solved self-consistently. As noted
in ref.[5-6](see also ref.[8]) the electric potential $\rho $  and  the
rhs of eq.(9) are related to fluctuations, which are  neglected  in
the conventional BO adiabatic approximation, in  the  presence  of
time $t$. Stated more precisely, the electric potential $\rho $  corresponds
to the first order nonadiabatic correction in the back reaction on
the $q$-modes from level transitions in the time dependent $\phi $ states,
which vanishes nevertheless in the  pure  adiabatic  limit. In  the
intrinsic time formalism(see below),on the other hand, $\rho $ turns  out
to be the dominant back reaction and cannot be dropped[8]. However,
the nonadiabatic corrections in the rhs of eq.(9)  is  clearly  of
higher order (in smallness) in a regime when the pure adiabaticity is 
weakly violated (c.f., ref [5] for an explicit calculation) and can 
be neglected safely in most of
the present discussion( these terms are important however for  the
energy conservation in the composite system).
\par
In  the  semiclassical  WKB  regime  of   the $q$-modes   one
substitutes for the effective wave function $\psi (q)$ the ansatz:
\par
$$
\psi  = \exp \left(i\int A_{i}dq^{i}\right) \psi _{eff} = 
\sigma\exp (i\int A_{i}dq^{i} +i{1\over \hbar } S )
\eqno(10)$$
\noindent in eq.(6), so that the effective  quasi-classical  
motion  of  the
$q$-modes is given by the Hamilton-Jacobi equation
\par
$$
{1\over 2M}{G^{ij}{\partial S \over\partial q^{i}}{\partial S \over
\partial q^{j}}} + MV +<H_{m}> +{1\over {2M}} \rho  =E
\eqno(11)$$
\noindent Here, $\sigma$(in eq(10)) is the WKB prefactor (the Van Vleck 
determinant) and the  intrinsic  (WKB)  time $\tau$ is  introduced  via  
the  vector field[3]
\par
$$
{d\over Nd\tau} = G^{ij}{\partial S \over\partial q^{i}}{\partial S 
\over \partial q^{j}} \rightarrow {dq^{i}\over Nd\tau} 
=G^{ij}P_{j}, P_{i}={\partial S \over \partial q^{i}}\eqno(12)$$

\noindent The  intrinsic  WKB  time $\tau$   parametrizes,   
upto   a   possible
reparametrization, the classical trajectories of  the $q$-modes  as
integral  curves   normal   to   the   level   surfaces   of   the
Hamilton-Jacobi function $S$. This is made  explicit  by  introducing
the lapse $N$ in eq.(12). The time $\tau$ thus corresponds, as it should, to
the Leibniz (Mach-Einstein) time[9], in contrast to  the  absolute
Newtonian time $t$ in eq.(2).
\par
The quantum evolution of the lighter state is thus  described
by (neglecting higher order fluctuations in eq.(9))
\par
$$
\left (H_{m}- <H_{m}>\right )\chi  = {i\hbar \over N}
\left ({d\over d\tau} -<\chi \mid {d\over d\tau}\chi >\right)\chi 
\eqno (13)$$

\noindent Note that eq.(13) (in fact eq.(9))  when  contracted  
with $\chi ^{*}$  is
satisfied identically. Each term of eq.(13) (eq.(9)) thus corresponds to
a pure fluctuation with zero mean. So far our discussion was perfectly general
and is valid for either of the two formalisms,with or  without  an
external time.The point of bifurcation ocurs as  one  proceeds  to
interpret eq.(13). Recall that our main goal in this paper is to emphasize
a {\em subtle} difference between the possible predictions of the two 
formalisms, which as it will turn out, can be amply illustrated even limiting
our discussion in the  regime when the pure adiabaticity is only weakly 
violated. Dropping of the RHS of eq.(9) is thus justified. In general,
the neglected term represents a higher order quantum gravitational correction
to the quantized matter evolution. The effects of this correction term on 
the matter evolution will be considered separately. We however emphasize that
although the term involving $\rho$ in eq.(11) encodes the effects of 
the higher order back reaction from the level transitions in the matter state, 
it nevertheless becomes important in the intrinsic formalism[8].
\par
In ordinary quantum  mechanics  with  an  a  priori  time $t$,
eq.(13) yields a unique, unambiguous interpretation(of
course, guided by the inputs from laboratory experiments). In  fact,
$\chi $ can be considered as the unique horizontal lift[10] of the ray 
$\tilde{\chi }$
satisfying  the  parallel  transport  law 
$<\chi \mid d/d\tau\mid \chi > =0$. Stated
otherwise, eq.(13) corresponds to the parallel transport  law  for
the Schr\"odinger equation
\par
$$
i\hbar {d\over dt} \chi _{s} = H_{m}\chi _{s}
\eqno(14)$$
\noindent where the Schr\"odinger state function is  given  
by $\chi _{s}=e^{-i\gamma }\tilde{\chi }$,  the
total phase $\gamma $ being  the  sum  of  the  dynamical  phase  and  the
adiabatic phase:$\gamma =\gamma _{d}+\gamma _{g}=\int <H_{m}>dt
-\int A_{i}dq^{i}$. Further,the time derivative
in the lhs of eq.(14) can  actually  be  thought  of  as  a  total
derivative:
\par
$$
{d\over dt}={\partial \over \partial t}+ {dq^{i}\over {Ndt}}
{\partial \over \partial q^i}
\eqno(15)$$
\noindent the second intrinsic derivative takes care  of  the  (adiabatic  )
fluctuation of the state $\chi _{s}$  over  its  mean  dynamical  evolution
(cf.,the rhs of eq.(13)).  Note  that  the BO  decomposition  (5)
realizes eq.(13) at a level when the evolution of  lighter  system
is {\it only} horizontal.The verticle (dynamical) component[1,10] of the
actual evolution in the Schr\"odinger state $\chi _{s}$ is recovered  through
the explicit time derivative from the external  time  $t$. Note  also
that the phases $\gamma _{d}$ and $\gamma _{g}$ are distinct,  
both  in  magnitude  and,
evidently, in their origin. $\gamma _{d}$ is purely dynamical and related  to
the external time $t$ (and to the Hamiltonian $H_m$), whereas the  latter  one
$\gamma _{g}$ is a geometric
contribution of the adiabatic $U(1)$ gauge group and independent  of
$t$. In   fact, $\gamma _{g}$   is   both   gauge   and   reparametrization
invariant[10]. It should however be {\it emphasized} that  a  nonzero 
$\gamma _{g}$
always  indicates  the  presence  of  a  small  scale  independent
evolution  in  the  quantum  state  over  the  dominant  dynamical
evolution associated with the mean energy $<H_{m}>$[8]. Note  that  the
above     treatment     is     strictly     (adiabatic)      gauge
invariant. Moreover, the total wave function $\Psi _{0}$ is also  independent
of the dynamical phase $\gamma _{d}$, because, as a consequence of the energy
conservation in the total system, the effective  wave  function $\psi $
also picks an equal dynamical phase $\gamma _{d}$, but for an opposite sign.
\par
As advocated already in ref.[8], the  small  scale  geometric
evolution could indeed be  exploited  to  introduce  a  meaningful
concept of intrinsic time in a (closed) dynamical  system  without
an external time. Below we show that this is,  in  fact,  the  only
reasonable choice, contrary to the recent claims[4-7].
\par
To begin with, let us first note that one could perhaps follow  formally
the steps similar to the above external time formalism even in the
closed  (gravity-matter  )  system[4-7]. This  is  because  of  the
explicit gauge invariance of the total wave function 
$\Psi _{0}=\psi \chi $ and the
almost identity like character of  eq.(13). One  could  thus  write
formally $\Psi _{0}=\tilde{\psi }\tilde{\chi }$, 
where $\psi =e^{i\gamma } \tilde{\psi }$ and 
$\chi =e^{-i\gamma } \tilde{\chi }, \gamma =\gamma _{d}+\gamma _{g}
=\int N<H_{m}>d\tau + \gamma _{g}$,
and then identify $\chi $  ,as  above,  as  the  Schr\"odinger  state  in
intrinsic time:
\par
$$
i\hbar {d\over Nd\tau} \chi  =H_{m}\chi 
\eqno(16)$$
\noindent This apparently fulfills the aim of  obtaining  the  semiclassical
Einstein equations with mean energy  as  back
reaction[3-7]. However, the above arguments  can  not  be  justified
rigorously. Note that the  split  of  the  total  phase $\gamma $  into  a
dynamical phase  and  a  geometric  phase  is  purely  formal  and could be 
misleading. In the absence of an  external  time,  the  expectation
value $<H_{m}>$ looses its distinguished  {\em dynamical} 
character  and  instead  gets
linked with the adiabatic gauge connection $A$. For, the total phase $\gamma $ in
this case {\it is geometric} and must  be  treated  as  a {\em single}  unit.
Explicitly, in the absence of time both $<H_{m}>$ and A gets related to
each   other   by   a   suitable   choice   of    the    adiabatic
gauge. Moreover, the reparametrization invariance of $\gamma $, rather  than
only of $\gamma _{g}$, is also evident. The validity of the  arguments  leading
to eq.(16) can thus be ascertained at best for a particular  gauge
choice:$<\chi \mid d/d\tau\mid \chi >=0$. There  are  however  other  equivalent
gauge
choices. One particularly  interesting  gauge  is $A=A_{i}dq^{i}=N<H_{m}>d\tau$
which together with eq.(13)  also  turns $\chi$ a Schr\"odinger  state
satisfying eq.(16) (total phase in $\chi $ in  this  case  is,  however,
zero in contrast to the former gauge with total phase $\gamma )$. However,
this gauge yields a different quasi  classical  equation  for  the
$q$-modes,viz.:
\par
$$
{1\over 2M} G^{ij}P_{i}P_{j} +MV +{1\over 2M} \rho  =E
\eqno(17)$$
\noindent which misses the zeroth order back reaction from the  mean  energy
$<H_{m}>$. The predictions of the theory thus  become  gauge  dependent
which is physically unacceptable. Further, there  is  as  such  no
reason to prefer one theory over another, at least at  this  level
of our analysis. An  extra  input  is  necessary  to  restrict  the
theory.Some aspects of this ambiguity has already  been  discussed
at  length  in  the  literature[11]. The   conclusions   drawn   in
refs.[4-7] are definitely motivated by the demand of obtaining the
semiclassical Einstein equations even in  a  quantum  cosmological
model.(However, the  semiclassical  Einstein  equations  with  mean
energy (-momentum tensor) as back reaction may  very  well  be  in
suspect(in the absence of any experimental clue) in the context of
quantum cosmology. In any case,the standard  semiclassical  theory
is expected to be unambiguous in the black hole back  ground  with
an asymptotic Minkowsky time). Our  approach[8]  differs  from  the
conventional ones in that we try to  interpret  eq.(13)  minimally
using only  the  demand  of  adiabatic  gauge  invariance  of  the
intrinsic formalism.
\par
To this end, one demands the theory to be gauge invariant,and
aims at expressing the quasi-classical equation (11) in  terms  of
gauge invariant quantities  eg.,the  adiabatic  field  strength $F
=dA,d$ being the exterior derivative in the $q$ space. To this  effect
we first compute $F$  in  the  gauge 
$A=N<H_{m}>d\tau=d\int^{q(\tau)}N<M_{m}>d\tau$. It
follows that $F=0$  and  hence  by  gauge  invariance, $F$  vanishes
globally. Consequently, the  flat  adiabatic  connection  $A_T$  corresponding 
to the total phase $\gamma$ of the state $\chi$ itself
vanishes for any  simply  connected  region  in  the $q$-space.(The
gravitational sector of the superspace is expected  to  be  simply
connected away from the big bang  singularity). The  triviality  of
the adiabatic gauge bundle  on  the $q$-space  tells  us  that  the total
adiabatic phase $\gamma$ (split formally as $\gamma_{d} + \gamma_{g}$ 
in eq.(16)) picked by the factored wave function $\chi$  (or $\psi$)  is
{\it necessarily} zero and consequently the associated  connection  must
not   contribute   in   the   gauge   invariant    equations    of
motion. Consequently,  the  mean  adiabatic  energy $<H_{m}>$( which appears now   
only as a part of the total connection $A_T$ ) must  be
unobservable in the intrinsic description. The
quasi-classical dynamics of the heavy system  thus remain {\it unaffected} by
the back reaction from the mean adiabatic energy  of  the  lighter
system in the intrinsic  description. The  back  reaction  is  then
determined only by the ( gauge invariant) electric potential 
$\rho $ which  happens  to  be
the first nonadiabatic correction on the adiabatic phase (zero  in
the   present   situation). The
quasi-classical equation of motion of the  heavy  system  {\it is}  thus
given by eq.(17)[8]:
\par
$$
{1\over 2M} G^{ij}P_{i}P_{j} +MV + {1\over 2M} \rho  =E
\eqno(18)$$
\par
The physical reason for the nonobservability of the  adiabatic
energy $<H_{m}>$  in  the  intrinsic  description may be stated thus[ 8]. The
measurement  of  energy  in  any  (localized)   dynamical   system
presupposes the existence of time and involves a definition of the
zero point. In the present model however  the  intrinsic  WKB  time
emerges only at a level when the quantized lighter system  follows
the quasi-classical $q$-modes adiabatically. The mutual  interactions
between the heavy and the lighter systems  induce,  on  the  other
hand, a gauge freedom in  the  composite  system,  which  in  turn
reflects a possible  arbitrariness in the choice of the zero point
in the measurement of energy of the (internal)  lighter  system. In
the case with an external time,the gauge freedom  is  harmless  as
the zero point in energy is  well  defined  from  the  outset  and
amounts  to  a suitable readjustment  in   the   scale   of   time $t$. More
importantly, the mean energy gets delinked from the adiabatic gauge
in the  presence  of  time  t. In  the  present  case, however, the
definition of intrinsic time is equivalent to a choice of gauge in
the Hilbert space of the lighter system, which in turn helps fixing
the zero point at the mean adiabatic energy  (the  only  available
value  of  energy  at  this  level), thus leading to a renormalization
in the matter Hamiltonian( $H_m \rightarrow (H_m - <H_m>)$).
Another  consequence  of   the
definition of time through a gauge choice is  that  it  makes  the
induced gauge interaction in the composite system trivial  at  the
adiabatic  level(  of  course, this  is  a   consequence   of   the
reparametrized nature of the intrinsic  time). The  zero  point  in
energy is thus realized globally so far as a single intrinsic time
variable suffices the description of the dynamics of the composite
system. Only the (gauge invariant)  energy  differences  over  the
mean adiabatic energy (ie.,the energy  fluctuations)  in  the
state $\chi $ thus have physical meaning in the intrinsic formalism.
\par
The Schr\"odinger equation (13) is now  interpreted  so  as  to
describe the time dependent nonstationary evolution of the lighter
state $\chi $, over the (nonobservable) adiabatic evolution described by
eq.(16).  This  is   achieved   by   the   method   of   dynamical
renormalization [12,13]. Explicitly, at the  adiabatic  level  the
state $\chi $ is realized as a stationary state $\chi _{o}$
\par
$$
H_{m}\chi _{0}=<H_{m}>\chi _{0}
$$
\noindent To capture the residual (fluctuating)  motion (in connection with   
the nonadiabatic geometric phase [8]) in $\chi $ in eq.(13), 
one  makes  a
unitary  transformation  by  writing $\bar{\chi } =U\chi$, 
where $U$ is the
interaction picture evolution operator: 
$U=\exp (-{i\hbar^{-1}}\int N\bar{H}_{m}d\bar{\tau})$. Here, $\bar{H}_{m}=
H_{m}-<H_{m}>=(dH_{m}/d\bar{\tau})d\bar{\tau}$,   
denotes   the   renormalized   (interaction)
Hamiltonian  and  the  time   variable $\bar{\tau}$   is   introduced   via
$id/d\bar{\tau}=id/d\tau-\hbar ^{-1}N<H_{m}>$, so that 
$d\chi /d\bar{\tau}=0$. The  actual time dependent evolution of the 
(transformed) fluctuating  state $\bar{\chi }$  is thus given by the 
(intrinsic) Schr\"odinger  equation
\par
$$
{i\hbar{d\over {d\bar {\tau}}}}\bar{\chi }=
N\bar{H}_{m}(\tau)\bar{\chi }(\bar{\tau})
\eqno(19)$$
\noindent with $\bar{\chi }(0)=\chi _{0}$. The (nonadiabatic) Pancharatnam 
connection [1,8] $A_{P}$ for the matter state $\chi$( eq.(13)) is now realized 
as the dynamical phase of the intrinsic equation (19):$A_{P}=
i<\phi\mid d\mid \phi >
=N\hbar^{-1}<\bar{\chi }\mid \bar{H}_{m}\mid \bar{\chi }>d\bar{\tau}=
N\Delta \epsilon d\bar{\tau}$, where $\chi = e^{-i\int A_{P}}\phi$ and $\phi
\in \cal P$, the projective ray space of the matter Hilbert space.
Moreover, $d$
stands for the corresponding ray space Fubini-Study exterior derivative  and 
$\Delta \epsilon $
is identified with the uncertainty in the  original  instantaneous
stationary   state $\chi _{0}$: $\Delta \epsilon =\sqrt{<\chi _{0}\mid 
(H_{m}-<H_{m}>)^{2}\mid \chi _{0}>}$[8,13].
We have thus completed a 
full circle
in  interpreting  eq.(13)  self  consistently  in   an   intrinsic
description.  In fact, eq.(19) is obtained as a verticle realization 
of the parallel transport law (13). This means in turn that the  
intrinsic geometric motion, in connection with the irreducible quantal 
fluctuations, in the state $\chi$ is realized as the dominant dynamical  
evolution for the state $\bar {\chi}$. Further, the relevant Hamiltonian
is now $\bar H_{m}$, instead of the original one $H_{m}$. 
The  present  derivation  of  the   intrinsic   time
Schrodinger eq.(19), being exact, not only  extends  our  previous
discussions[8] but also reflects clearly the role  played  by  the
irreducible  quantum   fluctuations   in   connection   with   the
nonadiabatic phase, in obtaining it. Note also the close similarity
of the two equations: the external time eq.(14) and the  intrinsic
time eq.(19)[13]. 
\par
To  sum  up,  eqs.(18)  and  (19)  constitute  the  two  main
equations in a  self  consistent  treatment  of  the BO  analysis
without an  external  time. This  set  of  equations  are  obtained
minimally from the exact set, eqs.(6) and (9) using the demand  of
adiabatic gauge invariance. This seems not only the correct set  of
equations  for  studying  the  semiclassical  limit  of  a  closed
gravity-matter system in quantum cosmology,  but  applications  of
this formalism even in ordinary quantum mechanics can not be ruled
out.Some  of   the   possible   approaches   have   already   been
discussed[8,13]. A more elaborate investigation is  under  way  and
will be reported elsewhere. In the framework of quantum  cosmology
(and semiclassical  gravity)  this  offers  an  insight  into  the
problem why the present value of the cosmological constant in  the
universe is negligibly  small;  the  reason  being,  as  envisaged
above,the nongravitating (nonobservable) nature of the mean vacuum
energy in the universe[8]. The intrinsic dynamical renormalization,
as presented above, seems to set the  renormalized  vacuum  energy
unambiguously to zero in  a  cosmological  background. The  present
discussion, however, is restricted  to  the  minisuperspace  of  a
homogeneous cosmology. The generalization of this study to  a  more
general superspace is an interesting problem for the future.
\par
We close with two remarks.
\par
\noindent 1.   Note  the  lapse  dependence  of  the  intrinsic  Schrodinger
equation (19). Though  the  freedom  in  the  adiabatic  gauge  is
utilized to define the zero point in energy in the lighter  system
thereby fixing the  characteristic  time  scale  for  the  quantum
evolution, the lapse in the  rhs  of  eq.(19)  indicates  a  still
existing residual freedom in its choice. As it is well  known  the
lapse dependence,in the case of a gravitational  background  is  a
consequence of  the  nonuniqueness  and  foliation  dependence  of
quantum mechanics( quantum field theory) in a curved space-time.In
the present formalism there is however an interesting choice which
maps eq.(19) to the extrinsic equation (14). Note that in the latter
equation the external time $t$ scales as $\epsilon ^{-1}$,
$\epsilon =<H_{m}>$  thus  fixing $N$
uniquely to $N=1$. In the intrinsic time  equation  one  however  has
$N\bar{\tau}\simeq (\Delta \epsilon )^{-1}= (\nu \epsilon )^{-1}$, 
$\nu <<1$. Thus with the choice $N=\nu ^{-1}$; $\bar{\tau}\simeq 
t\simeq \epsilon ^{-1}$. In this
fluctuation gauge, so to  speak,  the  residual  reparametrization
gauge freedom is removed, thus  allowing  one  to  (approximately)
identify the scaled eq.(19) with the external  time  eq.(14).  One
thus  reproduces  a  replica  of  the   original   external   time
Schr\"odinger equation  even in a smaller (intrinsic) time scale of  
the irreducible quantal fluctuations[13]. The dynamics of 
the heavy  system
in the two context however remain manifestly different.
\par
\noindent 2.   Both  the  gauge  invariant  electric  potential $\rho $  and  the
Pancharatnam  connection $A_{P}$  have  the   same   origin,viz:   the
uncertainty $\Delta \epsilon $ in the state $\chi $. Further,there is no violation
of the
energy conservation( unitarity)[6,8] in the total system  even  in
the present intrnisic time formalism.These follow from  the  exact
coupled eqs.(6) and (9) and the  observation  that  the  geometric
gauge connection consists, in general, of two parts: adiabatic  and
nonadiabatic(the  later  being  of  lower   order). The   adiabatic
component however vanishes in the  intrinsic  formalism. The  total
wave function $\Psi _{0}$ thus remains strictly gauge invariant in both the
formalisms: with or without time t. The non-unitarity  reported  in
refs.[3,7] is an artifact of the approximate  description  of  the
composite system.
\section*{Acknowledgement}
It is a  pleasure  to  thank  the High Energy Section, Iternational Centre for 
Theoretical Physics, Trieste, Italy for hospitality where part of the work is done. 
The author also thanks the Department of Science and Technology,Government of India 
for financial support( grant no. SP/S2/0-05/94).

\end{document}